\documentstyle[12pt]{article}
\def\1ad{\mbox{\normalsize $^1$}}
\def\2ad{\mbox{\normalsize $^2$}}
\def\3ad{\mbox{\normalsize $^3$}}
\def\4ad{\mbox{\normalsize $^4$}}
\def\5ad{\mbox{\normalsize $^5$}}
\def\6ad{\mbox{\normalsize $^6$}}
\def\7ad{\mbox{\normalsize $^7$}}
\def\8ad{\mbox{\normalsize $^8$}}
\def\makefront{\vspace*{1cm}\begin{center}
\def\newtitleline{\\ \vskip 5pt}
{\Large\bf\titleline}\\
\vskip 1truecm
{\large\bf\authors}\\
\vskip 5truemm
\addresses
\end{center}
\vskip 1truecm
{\bf Abstract:}
\abstracttext
\vskip 1truecm}
\def\half{{1 \over 2}}

\def\p{{\partial}}

\begin {document}
\def\titleline{
A New Approach to Superstring Field Theory
}
\def\authors{
Nathan Berkovits\\
IFT-P.034/99
}
\def\addresses{
Instituto de F\'\i sica Te\'orica, Universidade Estadual
Paulista\\
Rua Pamplona 145, 01405-900, S\~ao Paulo, SP, Brasil
}
\def\abstracttext{
I review the construction of an action for open superstring field
theory which does not suffer from the contact term problems of other
approaches. I also discuss a possible generalization of this action
for closed superstring field theory. (Talk presented
at $32^{nd}$ Ahrenshoop International Symposium on Elementary Particle
Theory in Buckow, Germany) }
\large
\makefront
\section {Problems with Conventional Approach}

The construction of a field theory action for the superstring is
an important problem since it may lead to information about
non-perturbative superstring theory which is unobtainable from the
on-shell perturbative S-matrix.
This information might be useful for understanding the non-perturbative
dualities of the superstring. Although there was much activity ten years
ago concerning a field theory action for the bosonic string, there
was not much progress on constructing a field theory action for the
superstring. For reasons described below,
there was no obvious generalization
of the bosonic string field theory action to the superstring.
Since the non-perturbative dualities of the superstring
are not expected to have any analog for the bosonic string, it is
not too surprising that not much was learned about these dualities
by studying bosonic string field theory.

The covariant string field theory action for the bosonic string is based on
a BRST operator $Q$ and a string field $A$. In Witten's approach to
open string field theory, the gauge-invariant action is
\cite{witbos}
\begin{equation}
S_{open}=\int [A Q A + A^3]
\label{openbos}
\end{equation}
where string fields are glued together at their
midpoint. This gluing prescription
preserves the length of the open string, so one
can fix the open string length in this approach. Witten's
gluing procedure can be generalized to closed strings by requiring
that two closed strings only interact when
half of one closed string overlaps with half of the other
closed string, so that the length of the resulting closed string is the
same as the length of each initial closed string.
Using this gluing procedure, gauge invariance
implies that
the field theory action is non-polynomial in $A$, i.e.
\cite{zw}
\begin{equation}
{S_{closed} =\int [A (c_0-\bar c_0) (Q+\bar Q) A + A^3 + ...]}
\label{closedbos}
\end{equation}
where $...$ includes
contributions from all orders in $A$. The factor $(c_0-\bar c_0)$
is necessary in
the kinetic term from ghost-number counting since $A$ carries ghost-number 2
and the closed string ghost-number anomaly is $+6$.
Because of the $(c_0-\bar c_0)$ factor, this closed string
action is gauge invariant under $\delta A=(Q+\bar Q)\Lambda$ only if
both $A$ and $\Lambda$ satisfy the constraints
$(b_0 -\bar b_0) A=(b_0 -\bar b_0)\Lambda=0$.

There is another covariant approach to bosonic string field theory in which
open strings are glued at their endpoints and closed strings are
glued at one point only \cite{siegel}\cite{hata}. This approach
is similar to light-cone string
field theory where the length of the string is the momentum in the
light-cone direction. In this covariantized light-cone
approach, the length of the string
becomes a free parameter which is conserved in interactions, i.e.
the length of the final string is equal to the sum of the lengths
of the two initial strings.
Although this free parameter can be gauged away
on-shell, it is unclear how to treat the infinities
caused by integration over this parameter in the functional integral.
Nevertheless, it is interesting to note that using this gluing prescription,
it is possible to construct a gauge-invariant
cubic closed string field theory action of the
form
\begin{equation}
S_{closed} =\int [A (c_0-\bar c_0) (Q+\bar Q) A + A^3].
\label{hata}
\end{equation}
As before,
$(b_0 -\bar b_0) A=0$ is required for gauge invariance.
One can also construct a gauge-invariant open string
field theory action using the endpoint gluing prescription, however, unlike
the action of (\ref{openbos}) (but like
the light-cone open string action), it is not cubic.

In generalizing these approaches to superstring field theory, the
main difficulty comes from the requirement that the string field carries
a definite ``picture''. Recall that each physical state of the superstring
is represented by an infinite number of BRST-invariant vertex operators
in the covariant RNS formalism \cite{fms}. 
To remove this infinite degeneracy, one
needs to require that that the vertex operator carries a definite picture,
identifying which modes of the $(\beta,\gamma)$ ghosts annihilate
the vertex operator.
For open superstring fields,
the most common choice is that all Neveu-Schwarz string fields carry
picture $-1$ and all Ramond string fields carry picture $-\half$.
For closed superstring fields, 
one has an analogous choice of (left,right) picture
corresponding to the choice of Neveu-Schwarz or Ramond in the (left,right)
sectors.

Since the total picture must equal $-2$ for open superstrings, the
obvious generalization of the action (\ref{openbos})
is
\cite{wsup}
\begin{equation}
S=\int [ A_{NS}
Q A_{NS} + A_R Q Y A_R  + Z ~A^3_{NS} +A_{NS} A_R A_R ]
\label{opensuper}
\end{equation}
where $Z=\{Q,\xi\}$ is the picture-raising operator of picture $+1$ and
$Y=c\p\xi e^{-2\phi}$ is the picture-lowering operator of picture $-1$.
However, as shown by Wendt
\cite{conw},
the action of (\ref{opensuper}) is
not gauge-invariant because of the contact-term divergences occuring when two
$Z$'s collide. Although one can choose other pictures for the string
field $A$ which change the relative factors of $Z$ and $Y$
\cite{pr},
there is
no choice for which the action is cubic and gauge-invariant
\cite{ya}
\cite{bigp}.
One way to make the action gauge-invariant
would be to introduce an infinite number of contact
terms to cancel the divergences coming from colliding $Z$'s.
However, the coefficients of these contact terms would have to
be infinite in the classical action since the divergences are present
already in tree-level amplitudes.
Note that an infinite number of divergent contact terms are also expected
in light-cone superstring field theory (either in the RNS or
Green-Schwarz formalisms) to cancel the divergences when
interaction points collide \cite{GK}.

For the closed superstring, the total left and right-moving pictures
must both equal $-2$, so the generalization of the actions of (\ref{closedbos})
and (\ref{hata})
is \cite{sen}
\begin{equation}
S_{closed} =\int [A_{NS,NS} (c_0-\bar c_0)
(Q+\bar Q) A_{NS,NS}+
\label{closedsuper}
\end{equation}
$$A_{NS,R} (c_0-\bar c_0)
(Q+\bar Q)\bar Y A_{NS,R} +
A_{R,NS} (c_0-\bar c_0)
(Q+\bar Q)Y A_{NS,R}
+ $$
$$
A_{R,R} (c_0-\bar c_0)
(Q+\bar Q)Y \bar Y A_{R,R}+  Z \bar Z A_{NS,NS}^3 + ...].$$
Since (\ref{closedsuper}) contains $Z$'s in the interactions,
this action naively suffers from the same contact-term divergences as
the action of (\ref{opensuper}). However, using the closed-string
gluing prescription of (\ref{closedbos}),
interaction points never collide so there is no problem.
But these contact-term divergences
are a problem using the covariantized light-cone gluing prescription
of (\ref{hata})
which allows colliding interaction points (as in light-cone string
field theory).

A problem with (\ref{closedsuper}) 
which exists using either gluing prescription
is that the picture-lowering operators
$Y$ and $\bar Y$ do not commute with $(b_0-\bar b_0)$. This
implies that the Ramond contribution to the action is not gauge-invariant
under $\delta A=(Q+\bar Q)\Lambda$
even at the quadratic level. Such a problem with the Ramond sector is
not surprising since the Type IIB Ramond-Ramond sector
contains a massless chiral
four-form state for which it is extremely difficult to construct a kinetic
action.

\section{ New Approach for the Open Superstring} 

In these proceedings, a new approach to superstring field theory will
be proposed which uses two BRST-like operators instead of just one.
This approach is based on the fact that any critical N=1 superconformal
field theory (such as the ten-dimensional
superstring) can be described by a critical
N=2 superconformal field theory. This
N=1 $\to$ N=2 embedding was described
in reference \cite{UST} where it was shown that any physical N=1 vertex
operator can be represented by a physical N=2 vertex operator,
and the scattering amplitudes coincide
using either the N=1 or N=2 prescriptions for computation.
Furthermore, it was shown in reference \cite{top} that, after twisting,
N=2 physical vertex operators
and scattering amplitudes can be computed
without introducing N=2 ghosts. 

This ghost-free prescription
was called
an N=4 topological prescription since
it uses the (small) N=4 superconformal generators which can be constructed
from any set of critical N=2 superconformal generators.
The four fermionic
N=4 generators will be labeled as ($G^+, G^-, \tilde G^+, \tilde G^-$)
where $G^+$ and $G^-$ are the original fermionic
N=2 generators, $\tilde G^+ = [e^{\int J}, G^-]$,
$\tilde G^-= [e^{-\int J}, G^+]$, and ($e^{\int J}, J, e^{-\int J})$ are
the SU(2) generators constructed from the original
U(1) generator $J$.
After twisting, $(G^+,\tilde G^+)$ carry conformal
weight $+1$ and $(G^-,\tilde G^-)$ carry conformal weight $+2$.

There are three critical N=2 superconformal field theories which
will be relevant here. The first is the self-dual string which describes
self-dual Yang-Mills (open) or self-dual gravity (closed) in $D=(2,2)$ \cite
{oog}. The left-moving worldsheet fields of the self-dual string consist of 
$(x^{\pm a},\psi^{\pm a})$ where $a=1$ to 2.
For the self-dual string, the left-moving
N=4 fermionic generators are 
\begin{equation}
G^+ = \p x^{-a}\psi^{+a},~ \tilde G^+ = \p x^{+a} \psi^{+a},~
G^- = \p x^{+a}\psi^{-a},~ \tilde G^- = \p x^{-a} \psi^{-a}.
\label{selfdual}
\end{equation}

A second critical N=2 superconformal field theory is given by the
N=2 embedding of the RNS superstring. The worldsheet fields are
the usual RNS worldsheet variables and the N=4
fermionic generators are \cite{twisted}\cite{top}
\begin{equation}
G^+ = j_{BRST},~ \tilde G^+ = \eta,~
G^- = b,~ \tilde G^- = \{Q, b\xi\} = b Z + \xi L
\label{RNS}
\end{equation}
where $Q=\int j_{BRST}$ is the standard BRST charge of the N=1 superstring,
$\eta$ and $\xi$ come from bosonizing the $(\beta,\gamma)$ ghosts as
$(\beta = \p\xi e^{-\phi}, \gamma= \eta e^\phi)$ \cite{fms}, $Z$ is
the picture-raising operator, and $L$ is the RNS stress-tensor.

A third critical N=2 superconformal field theory is given by a modified
version of the 
Green-Schwarz superstring which describes in $D=4$ superspace the
superstring compactified on a six-dimensional manifold \cite{four}. 
The left-moving worldsheet fields of this superstring
consist of $[x^m, \theta^\alpha,$
$ \bar\theta^{\dot\alpha},$
$ p_\alpha,$
$\bar p_{\dot\alpha},$
$ \rho]$ plus an N=2 c=9 superconformal field theory
which describes the compactification manifold. $p_\alpha$
and $\bar p_{\dot\alpha}$ are the conjugate momenta to the superspace
variables $\theta^\alpha$ and $\bar\theta^{\dot\alpha}$, $\rho$
is a chiral boson, $m=0$ to 3, and $\alpha,\dot\alpha=1$ to 2. 
As discussed in \cite {four}, this superstring
is related to the RNS superstring by a field-redefinition of the worldsheet
variables. For this version of
the superstring, the four fermionic generators are 
\begin {equation}
G^+ = d^\alpha d_{\alpha} e^\rho + G^+_C, \quad
\tilde G^+ = [e^{\int J}, G^-],
\label{GS}
\end {equation}
$$G^- = \bar d^{\dot\alpha} \bar d_{\dot\alpha} e^{-\rho} + G^-_C,\quad
\tilde G^- = [e^{-\int J}, G^+],$$
where $d_\alpha=p_\alpha +i\bar\theta^{\dot\alpha}\p x_{\alpha\dot\alpha}$,
$\bar d_{\dot\alpha}=\bar p_{\dot\alpha} +i\theta^{\alpha}\p 
x_{\alpha\dot\alpha}$,
$J=\p\rho + J_C$, and
$[T_C,$
$ G^+_C,$
$ G^-_C,$
$ J_C]$ are the c=9 N=2 superconformal generators
representing the compactification.

In N=4 topological language, the on-shell condition for an N=2 open
string vertex operator
$V$ is \cite {top}
\begin{equation}
G^+_0 \tilde G^+_0 V=0
\label{eom}
\end{equation} where $G^+_0$ and $\tilde G^+_0$ are the zero
modes of two of the four N=4 superconformal generators. This linearized
equation of motion is invariant under the linearized gauge invariances
$\delta V =G^+_0 \Lambda + \tilde G^+_0 \tilde\Lambda$. 
Note that after twisting, the hermitian conjugate of $G^+$ is $\tilde G^+$ and
$V$ is a hermitian string field.
For the N=2
superconformal field theory representing the RNS superstring, the
N=2 vertex operator $V$ is related to the usual RNS vertex operator $A$
by $V=\xi_0 A$ (or equivalently, $\eta_0 V = A$). Since (\ref{RNS}) implies
that $G^+_0=Q$ and $\tilde G^+_0= \eta_0$,
the on-shell condition of (\ref{eom}) is equivalent to the usual RNS
on-shell condition $Q A=0$.

So the kinetic action $\int A Q A$ is naturally replaced with
the action \cite{sft}
\begin{equation}
S_{kinetic}=\int V G^+_0 \tilde G^+_0 V.
\label{kin}
\end{equation}
Under the U(1) of the N=2 algebra,
$G^+$ and $\tilde G^+$ carry charge $+1$ and $V$ is neutral, so
the open string action violates U(1) charge by $+2$ as expected
for a twisted critical N=2 superconformal field theory.
This kinetic action for N=2 superconformal field theories
is naturally extended to the non-linear action \cite{sft}
\begin{equation}
S_{open} = \int [(g^{-1}G^+_0 g )(g^{-1}\tilde G^+_0 g) + \int dt
(g^{-1} G^+_0 g) (g^{-1}\tilde G^+_0 g)(g^{-1} \p_t g)]
\label{opentwo}
\end{equation}
where $g= e^V$ and the open string fields $V$ are glued together
using Witten's midpoint prescription.
As in the Wess-Zumino-Witten action, this action has the non-linear
gauge invariance
\begin{equation}
\delta (e^V)= G^+_0(\Lambda) e^V + e^V \tilde G^+_0(\tilde\Lambda)
\label{opengauge}
\end{equation}
which replaces the linearized gauge invariances of (\ref{eom}).

When the N=2 superconformal field theory is
chosen to be the open self-dual string, it
is easy to show that (\ref{opentwo}) correctly reproduces the field
theory action for self-dual Yang-Mills.
When the N=2 superconformal field theory is chosen to be
the modified Green-Schwarz open superstring of (\ref{GS}),
the action of (\ref{opentwo}) provides
an open superstring field theory action with manifest
four-dimensional spacetime supersymmetry which does not suffer
from the contact-term divergences of all other open superstring actions.
For the uncompactified superstring, one can 
easily show \cite{sft} that the 
massless contribution to (\ref{opentwo}) correctly reproduces the
action for $D=10$ super-Yang-Mills 
written in terms of N=1 $D=4$ superfields \cite{marcus}.
Note that the gauge invariance of (\ref{opengauge}) is
reminiscent of the gauge transformation of the super-Yang-Mills
prepotential in N=1 D=4 superspace.

\section{New Approach for the Closed Superstring}

In this section, some preliminary results are presented for the action for
closed superstring field theory. The kinetic action of the open superstring
can be generalized to the following closed string kinetic action:
\begin{equation}
S=\int V (J^{++}_0+\bar J^{++}_0)
(G^+_0 +\bar G^+_0)(\tilde G^+_0 +\bar{\tilde G}^+_0)V 
\label{kinsuper}
\end{equation}
where $J^{++}=e^{\int J}$ is one of the SU(2) generators of the N=4 algebra.
This action correctly reproduces the kinetic action for self-dual gravity
when the N=2 superconformal
field theory is chosen to be the closed self-dual string.
Furthermore, when the N=2 superconformal
field theory is chosen to be the closed RNS superstring of (\ref{RNS})
and when
$V$ is a (NS,NS) string field satisfying $(G^-_0 -\bar G^-_0) V=0$
(which is the N=4 topological analog of
$(b_0-\bar b_0)A=0$),
the action of
(\ref{kinsuper}) reproduces the (NS,NS) contribution to
the kinetic action of (\ref{closedbos}). This is easy to see since
$(G^-_0 -\bar G^-_0)V=0$ implies that $V=(G^-_0 -\bar G^-_0) W$ for
some $W$, so the action of (\ref{kinsuper}) is
$$S=\int V (J^{++}_0+\bar J^{++}_0)
(G^+_0 +\bar G^+_0)(\tilde G^+_0 +\bar{\tilde G}^+_0) 
(G^-_0 -\bar G^-_0)W$$
$$=\int V (\tilde G^{+}_0-\bar {\tilde G}^{+}_0)
(G^+_0 +\bar G^+_0)(\tilde G^+_0 +\bar{\tilde G}^+_0) 
W
=2\int V \eta_0\bar \eta_0 (Q+\bar Q)W,$$
$$=2\int V \eta_0\bar \eta_0 (Q+\bar Q)(c_0 -\bar c_0)V,$$
which is the usual RNS kinetic action if one defines $V=\xi_0\bar\xi_0 A$
(or equivalently, $A=\eta_0\bar\eta_0 V$).

It has not yet been possible to use (\ref{kinsuper}) to
derive the Ramond-Ramond contribution to the kinetic action, or to
derive the modified Green-Schwarz version of the kinetic action (which
would be manifestly spacetime supersymmetric).
Note that the Ramond-Ramond contribution to the kinetic
action was computed in reference
\cite{emdual},
but using a different method \cite
{bigp}.
One problem with (\ref{kinsuper}) is that the constraint
$(G^-_0 - \bar G^-_0) V=0$ is not hermitian since the hermitian conjugate
of $(G^-_0 -\bar G^-_0)$ is $(\tilde G^-_0 - \bar{\tilde G}^-_0)$.
Although gauge invariance of (\ref{kinsuper}) does not
require that $V$ satisfies
$(G^-_0 - \bar G^-_0) V=0$, it appears that such a constraint is necessary
for introducing interactions.

For generalizing (\ref{kinsuper})
to include interactions, one can adopt either of the
two
gluing prescriptions. The covariantized light-cone gluing
prescription has the advantage that there is a Jacobi-like identity
satisfied by switching the order of gluing three strings, i.e.
$(A(BC)) = ((BA)C) +((AC)B)$ where $A,B$ and $C$
are the three strings and $()$ denotes the gluing procedure.
Using this gluing prescription and ignoring the hermiticity
problem mentioned above, one can define
the gauge-invariant non-linear action as
\begin{equation}
S=\int (~ V (J^{++}_0+\bar J^{++}_0)
(G^+_0 +\bar G^+_0)(\tilde G^+_0 +\bar{\tilde G}^+_0)V +
 [V,(G^+_0 +\bar G^+_0)V]
(G^+_0 +\bar G^+_0) V~) 
\label{closedtwo}
\end{equation}
where $[A,B]$ is defined by gluing
$(\tilde G^+_0 - \bar {\tilde G}^+_0)A$ with
$(\tilde G^+_0 - \bar {\tilde G}^+_0)B$ (which is anti-symmetric
in $A$ and $B$).
If one ignores hermiticity problems, it
appears to be also possible to construct a non-linear version 
of (\ref{kinsuper})
using the Witten-like gluing prescription.

If $V$ satisfies the constraint 
$(G^-_0 -\bar G^-_0)V=0$, then  (\ref{closedtwo}) is
invariant under 
$$\delta V =
(G^+_0+\bar G^+_0)\Lambda
+ (\tilde G^+_0-\bar {\tilde G}^+_0)\tilde\Lambda + $$
$$ 
(\tilde G^+_0+\bar {\tilde G}^+_0)\Omega +(G^-_0-\bar G^-_0)((\tilde
G^+_0 + \bar{\tilde G}^+_0)\Omega~
(G^+ + \bar G^+)V).$$
When the N=2 superconformal field
theory is chosen to be the closed self-dual string,
(\ref{closedtwo}) reproduces the correct
cubic action for self-dual gravity. And when the N=2 superconformal
field theory is chosen to be the RNS superstring, (\ref{closedtwo})
reproduces the correct (NS,NS) contribution.

It is easy to show that the covariantized light-cone gluing prescription
implies that
$[A,[B,C]]+[B,[C,A]]+[C,[A,B]]=0,$
so the interaction term can be interpreted as
the large N limit of an open string
interaction term
where $[,]$ for open strings
is the usual commutator coming from the U(N) Chan-Paton factors.
This is not so surprising since
self-dual gravity can be interpreted as the large N limit of
self-dual Yang-Mills, suggesting that the closed self-dual
string can be interpreted as the large N limit of the open
self-dual string \cite{jev}. If such an interpretation also holds for
the N=2 superconformal field theory
representing the ten-dimensional superstring, it might
shed light on the duality relating N=4 super-Yang-Mills with the Type IIB
superstring \cite {mal}.

{\bf Acknowledgements:} I would like to thank Warren Siegel, Cumrun
Vafa and Barton Zwiebach for their collaboration on parts of this work
and CNPq grant 300256/94-9 for partial financial support.


\end{document}